\documentclass[aps,preprint]{revtex4}%
\usepackage{amsfonts}
\usepackage{amsmath}
\usepackage{amssymb}
\usepackage{graphicx}%
\begin{document}
\title{{\LARGE ON THE MACHIAN ORIGIN OF INERTIA}}
\author{Marcelo Samuel Berman$^{1}$}
\affiliation{$^{1}$Instituto Albert Einstein/Latinamerica - Av. Candido Hartmann, 575 -
\ \# 17}
\affiliation{80730-440 - Curitiba - PR - Brazil - email: msberman@institutoalberteinstein.org}
\date{First Manuscript: 23 September, 2006; Last: 04 September, 2008}

\begin{abstract}
We examine Sciama's inertia theory: we generalise it, by combining rotation
and expansion in one unique model, we find the angular speed of the Universe,
and we stress that the theory is zero-total-energy valued. We compare with
other theories of the same null energy background.

We determine the numerical value of a constant which appears in the Machian
inertial force expression devised by Graneau and Graneau[2], by introducing
the above angular speed. We point out that this last theory is not restricted
to Newtonian physics as those authors stated but is, in fact, compatible with
other cosmological and gravitational theories. An argument by Berry[7] is
shown in order to "derive" Brans-Dicke relation in the present context.



\end{abstract}
\maketitle

\begin{center}
{\LARGE ON THE MACHIAN ORIGIN OF INERTIA}

\bigskip

Marcelo Samuel Berman
\end{center}

\bigskip

{\large Section I - Introduction}

The advancement of scientific knowledge has produced, in the aftermath of
Newtonian theory, the surge of General Relativity and alternative versions of
it. On a different line of thought, Sciama[1] has put forward a gravity theory
based on an electrodynamical analogy.

\bigskip

The research on the origin of Inertia is a problem that involved passionately
a number of physicists especially from the crisis of Classical Physics and
from the birth of General Relativity. Einstein himself underlined the
importance of his meditation on this argument in the development of his theory
of gravitation. Moreover he formulated precisely his reflections on inertia
origin in the Mach's Principle - that in its simpler form says that inertia
properties of matter are determined in some manner by the other bodies of the
Universe. Even though with the development of General Relativity Einstein
rejected explicitly his first considerations on Inertia, the Principle
represented, in a compact form, a research project that guided many scientists
in the development of gravitational theories alternative to General Relativity
(for example Brans-Dike Theory) - with the following cosmological implications
- and the reformulations of Classical Mechanics based on Mach's reflections on
Inertia (see for example the model proposed by Shr\"{o}dinger).

This short introduction has only the simple task to set in his historical
background the work submitted by the present author, that moves on these lines
of thoughts. This letter, has a double structure: in the beginning part we
present changes in the Sciama cosmological model (linked to gravitational
theories alternative to General Relativity) with the introduction of the
expansion and rotation of the Universe, and in the second part of the article
we use the results just obtained (the angular velocity of the Universe) in the
calculation of a constant of Graneau and Graneau theory. Moreover at the end
of this letter, I present in a simple form an argument proposed by Berry,
where the Machian analysis of inertia and the alternative cosmological model
(like Sciama and Brans-Dike model) shows some common aspects.

The suggestion of a Universe in rotation - certainly the most interesting and
the most problematic - is better understood in my prior papers ([3]-[8]), so
leaving a small gap in the presentation. Also the Sciama cosmological model is
presented in a manner that I myself consider a bit concise.

Another critical point, a key argument to link Machian Universe to Graneau and
Graneau Theory, is the extremely concise - but I think extremely interesting -
presentation of my interpretation of the Mach Principle as the mathematical
representation of the zero-total-energy of the Universe and the consequent
description of the Pioneer effect [3] - [8].

This letter also creates interesting links among research programs, like the
cosmological model based on Field Theory and Newtonian Machian Physics, that
are often considered in competition.

I consider this letter as an original contribution to the debate on the origin
of inertia and the connected argument of the cosmological model: so I think
the arguments proposed in this manuscript are very suitable.

\bigskip

\bigskip In Section II, it is shown that Sciama's paper can be generalised, by
including, in the same model, a rotating and expanding Universe; this also
implies that the total energy density of the Universe is null, at each point
of space; on summing for the whole Universe, we obtain zero-total energy. A
possible angular speed proportional to the inverse of the scale-factor is
found, which is of the same form of that found by Berman(2007; 2007a; 2007b;
2008; 2008a; 2008b).

\bigskip

In Section III, we review a theory with the otherwise "arbitrary" constant
\ \ $B$\ \ in the inertia formula, by Graneau and Graneau[2]. The numerical
value was in fact, found by us, in accordance with a possible rotation of the
Universe. In Section IV, a Machian argument by Berry[9] is shown to be related
with Sciama's theory.

\bigskip

{\large \bigskip Section II - Sciama's Inertia Model}

\bigskip In order to fulfill a theoretical need for accounting the inertia
properties of matter, Sciama supposes that gravitation is analogous to
electrodynamics. The working hypothesis is that inertial and gravitational
forces cancel each other at any point of space, so that the total field is
null. We combine first the calculations of expansion and rotation of the
Universe; each one was treated by Sciama, isolated.

\bigskip

For a rest particle, the "electric" potential contribution from the whole
Universe, as observed in time \ \ "$t$"\ \ \ is given by:

\bigskip

$\Phi=-\int\frac{\rho}{r}dV$%
\ $\ \ \ \ \ \ \ \ \ \ \ \ \ \ \ \ \ \ \ \ \ \ \ \ \ \ \ $\ \ \ \ \ \ \ . \ \ \ \ \ \ \ \ \ \ \ \ \ \ \ \ \ \ \ \ \ \ \ \ \ \ \ \ \ \ \ \ \ \ \ \ \ \ \ \ \ \ \ \ \ \ \ \ \ \ \ \ (1)

\bigskip

If the density is uniform, we have from (1):

\bigskip

$\Phi\cong-2\pi\rho c^{2}\tau^{2}$
\ \ \ \ \ \ \ \ \ \ \ \ \ \ \ \ \ \ \ \ \ \ \ \ \ \ \ \ \ \ \ \ , \ \ \ \ \ \ \ \ \ \ \ \ \ \ \ \ \ \ \ \ \ \ \ \ \ \ \ \ \ \ \ \ \ \ \ \ \ \ \ \ \ \ \ \ \ \ \ \ \ \ \ \ \ (2)

\bigskip

while, by symmetry, the "vector" potential is null:

\bigskip

$\vec{A}=0$ \ \ \ \ \ \ \ \ \ \ \ \ \ \ \ \ \ \ \ \ \ \ \ \ \ \ \ . \ \ \ \ \ \ \ \ \ \ \ \ \ \ \ \ \ \ \ \ \ \ \ \ \ \ \ \ \ \ \ \ \ \ \ \ \ \ \ \ \ \ \ \ \ \ \ \ \ \ \ \ \ \ \ \ \ \ \ \ \ \ \ \ \ \ \ \ \ \ \ \ (3)

\bigskip

In formula (2), \ $\tau$\ \ is associated with \ $H^{-1}$\ , where
\ $H$\ \ stands for Hubble's parameter.

\bigskip

If the particle moves with a linear velocity \ $\vec{v}$\ \ , and we add
Hubble's expansion, its total velocity is \ $\left[  \vec{v}+\vec{r}H\right]
$\ \ \ , at any point in the Universe. In the first approximation, $\Phi
$\ keeps approximately the same form as above. However, the vector potential
\ $\vec{A}$\ \ , when \ $\vec{v}$\ \ does not depend on \ $\vec{r}$\ \ , will
be given by:

\bigskip

$\vec{A}\cong-\int\frac{\rho\vec{v}}{cr}dV\cong\frac{\Phi\vec{v}(t)}{c}%
$\ $\ \ \ \ \ \ \ \ \ \ \ \ \ \ \ \ \ \ \ \ \ \ \ $\ \ \ \ \ \ \ . \ \ \ \ \ \ \ \ \ \ \ \ \ \ \ \ \ \ \ \ \ \ \ \ \ \ \ \ \ \ \ \ \ \ \ \ \ \ \ \ \ \ \ \ \ (4)\ 

\bigskip

\bigskip The "electric" field is given by the usual electromagnetic formula:

\bigskip$\vec{E}=-\triangledown\Phi-\frac{1}{c}\frac{\partial\vec{A}}{\partial
t}\cong-\frac{1}{c^{2}}\Phi\frac{\partial\vec{v}}{\partial t}$
\ \ \ \ \ \ \ \ \ \ \ \ \ \ \ \ \ \ \ \ . \ \ \ \ \ \ \ \ \ \ \ \ \ \ \ \ \ \ \ \ \ \ \ \ \ \ \ \ \ \ \ \ \ \ \ \ \ \ \ \ \ \ \ \ \ (5)

\bigskip

\bigskip The above is called in this case, a gravitoelectric field.

\bigskip

On the other hand, we would have a null gravitomagnetic field,

\bigskip

$\vec{B}=\triangledown$ $\times$ $\vec{A}\cong0$
\ \ \ \ \ \ \ \ \ \ \ \ \ \ \ \ \ \ \ \ \ . \ \ \ \ \ \ \ \ \ \ \ \ \ \ \ \ \ \ \ \ \ \ \ \ \ \ \ \ \ \ \ \ \ \ \ \ \ \ \ \ \ \ \ \ \ \ \ \ \ \ \ \ \ \ \ \ \ \ \ \ \ \ \ \ \ (6)

\bigskip

Now, let us have a body of mass \ $M$\ \ , placed in the Universe. In the rest
frame of a particle at rest, the total "electric" field will be given by:

\bigskip

$E_{TOT}\cong-\frac{1}{c^{2}}\Phi\frac{d\vec{v}}{dt}+\left[  \frac{M}{r^{2}%
}+\frac{\phi}{c^{2}}\frac{d\vec{v}}{dt}\right]  $
\ \ \ \ \ \ \ \ \ \ \ \ \ \ \ . \ \ \ \ \ \ \ \ \ \ \ \ \ \ \ \ \ \ \ \ \ \ \ \ \ \ \ \ \ \ \ \ \ \ \ \ \ \ \ \ \ \ \ \ \ \ \ \ \ (7)

\bigskip

In the above, \ \ $\phi$\ \ is the usual potential of the body with mass
$M$\ , on a test particle, i.e.

\bigskip

$\phi=$\ \ $-\frac{M}{r}$\ \ \ \ \ \ \ \ \ \ \ \ \ \ \ \ \ \ \ \ \ \ .\ \ \ \ \ \ \ \ \ \ \ \ \ \ \ \ \ \ \ \ \ \ \ \ \ \ \ \ \ \ \ \ \ \ \ \ \ \ \ \ \ \ \ \ \ \ \ \ \ \ \ \ \ \ \ \ \ \ \ \ \ \ \ \ \ \ \ \ \ \ \ \ \ \ \ \ \ (8)

Now, we write the total field as:

\bigskip

$E_{TOT}\cong\frac{M}{r^{2}}+\frac{1}{c^{2}}\left[  \phi-\Phi\right]  \vec{a}$
\ \ \ \ \ \ \ \ \ \ \ \ \ \ \ \ \ \ \ \ \ \ \ \ \ \ \ , \ \ \ \ \ \ \ \ \ \ \ \ \ \ \ \ \ \ \ \ \ \ \ \ \ \ \ \ \ \ \ \ \ \ \ \ \ \ \ \ \ \ \ \ \ (9)

\bigskip

where \ $\vec{a}$\ \ is the total acceleration (of the Universe plus the
body), relative to the test particle. Put it in other frame:\ the particle
accelerates towards the rest body, relative to the whole Universe.

\bigskip

\bigskip As this theory is of the electromagnetic type, it is a linear one. We
superimpose, then, the cases of \ radial \ expanding, and rotating pictures.

In the first place, consider a non-rotating Universe; now we set a reference
frame with origin at the body; in relativistic units, near the origin, we
shall have the scalar and vector potentials given by:

\bigskip

$\vec{A}=0$ \ \ \ \ \ \ \ \ \ \ \ \ \ \ \ \ \ \ \ \ \ \ \ , \ \ \ \ \ \ \ \ \ \ \ \ \ \ \ \ \ \ \ \ \ \ \ \ \ \ \ \ \ \ \ \ \ \ \ \ \ \ \ \ \ \ \ \ \ \ \ \ \ \ \ \ \ \ \ \ \ \ \ \ \ \ \ \ \ \ \ \ \ \ \ \ \ \ (10)

\bigskip

and,

\bigskip

$\Phi\cong-1$ \ \ \ \ \ \ \ \ \ \ \ \ \ \ \ \ \ \ \ \ \ . \ \ \ \ \ \ \ \ \ \ \ \ \ \ \ \ \ \ \ \ \ \ \ \ \ \ \ \ \ \ \ \ \ \ \ \ \ \ \ \ \ \ \ \ \ \ \ \ \ \ \ \ \ \ \ \ \ \ \ \ \ \ \ \ \ \ \ \ \ \ \ \ \ (11)

\bigskip

In a rotating Universe, however, if the axis of rotation is the Z ,
then\ \ \ near the origin, we shall have:

\bigskip

$A_{x}\cong\omega y$ \ \ \ \ \ \ \ \ \ ,

\bigskip

$A_{y}\cong-\omega x$ \ \ \ \ \ \ \ \ \ ,

\bigskip\ \ \ \ \ \ \ \ \ \ \ \ \ \ \ \ \ \ \ \ \ \ \ \ \ \ \ \ \ \ \ \ \ \ \ \ \ \ \ \ \ \ \ \ \ \ \ \ \ \ \ \ \ \ \ \ \ \ \ \ \ \ \ \ \ \ \ \ \ \ \ \ \ \ \ \ \ \ \ \ \ \ \ \ \ \ \ \ \ \ \ \ \ \ \ \ \ \ \ \ \ \ \ \ \ \ \ \ \ \ \ \ (12)

$A_{z}\cong0$ \ \ \ \ \ \ \ \ \ ,

\bigskip

$\Phi^{rot}\cong-\left[  1+\omega^{2}r^{2}\right]  ^{1/2}$ \ \ \ \ \ \ \ \ \ \ \ \ \ \ \ \ \ \ .

\bigskip

\bigskip In this case, the total field, will be given by:

\bigskip

$\vec{E}_{TOT}^{rot}=-\triangledown\Phi^{rot}-\frac{\partial\vec{A}}{\partial
t}\cong-\frac{\omega^{2}r}{\left[  1+\omega^{2}r^{2}\right]  ^{1/2}}+\frac
{M}{r^{2}}$ \ \ \ \ \ \ \ \ \ \ \ \ \ . \ \ \ \ \ \ \ \ \ \ \ \ \ \ \ \ \ \ \ \ \ \ \ \ \ \ \ \ \ \ \ \ \ (13)

\bigskip

\bigskip Now, we equate, on the test particle, inertia to gravitation, one
balancing the other, so that the total field is zero, where this total is not
the above only but added to \ $\vec{E}_{TOT}$\ \ , which equate to zero:

\bigskip

$-\frac{\omega^{2}r}{\left[  1+\omega^{2}r^{2}\right]  ^{1/2}}+\frac{2M}%
{r^{2}}-2\pi\rho H^{-2}\cong0$ \ \ \ \ \ \ \ \ \ \ \ \ \ \ \ \ \ \ \ \ . \ \ \ \ \ \ \ \ \ \ \ \ \ \ \ \ \ \ \ \ \ \ \ \ \ \ \ \ \ \ \ \ \ \ \ \ \ \ \ \ (14)

\bigskip

A particular solution, is composed by the two equalities below, whose sum
retrieves the above one:

\bigskip

$\frac{M}{r^{2}}\cong2\pi\rho H^{-2}a\ \ \ \ \ \ $
\ \ \ \ \ \ \ \ \ \ \ \ \ \ \ \ \ \ \ \ . \ \ \ \ \ \ \ \ \ \ \ \ \ \ \ \ \ \ \ \ \ \ \ \ \ \ \ \ \ \ \ \ \ \ \ \ \ \ \ \ \ \ \ \ \ \ \ \ \ \ \ \ \ \ \ \ \ (15)

\bigskip

(However, the Newtonian acceleration, in the above, is given by \ $a=-\frac
{M}{r^{2}}$\ ).\ 

$\frac{\omega^{2}r}{\left[  1+\omega^{2}r^{2}\right]  ^{1/2}}\cong\frac
{M}{r^{2}}$ \ \ \ \ \ \ \ \ \ \ \ \ \ \ \ \ \ \ \ \ \ \ \ \ \ \ \ \ \ . \ \ \ \ \ \ \ \ \ \ \ \ \ \ \ \ \ \ \ \ \ \ \ \ \ \ \ \ \ \ \ \ \ \ \ \ \ \ \ \ \ \ \ \ \ \ \ \ \ \ \ \ \ (16)

\bigskip

Equality (15) is satisfied by the Whitrow-Randall expression,

\bigskip

$G\rho H^{-2}\cong1$ \ \ \ \ \ \ \ \ \ \ \ \ \ \ \ \ \ \ \ \ \ \ \ \ \ \ \ . \ \ \ \ \ \ \ \ \ \ \ \ \ \ \ \ \ \ \ \ \ \ \ \ \ \ \ \ \ \ \ \ \ \ \ \ \ \ \ \ \ \ \ \ \ \ \ \ \ \ \ \ \ \ \ \ \ \ \ \ \ \ (17)

\bigskip

The above is equivalent, because \ $\rho\cong\frac{M}{\frac{4}{3}\pi r^{3}}%
$\ \ , to the Brans-Dicke form,

\bigskip

$\frac{GM}{r}=\gamma\sim1$\ \ \ \ \ \ \ \ \ \ \ \ \ \ \ \ \ \ \ \ \ \ \ \ . \ \ \ \ \ \ \ \ \ \ \ \ \ \ \ \ \ \ \ \ \ \ \ \ \ \ \ \ \ \ \ \ \ \ \ \ \ \ \ \ \ \ \ \ \ \ \ \ \ \ \ \ \ \ \ \ \ \ \ \ \ \ \ \ \ (18)

\bigskip

Notice that with the above, mass and radius are directly proportional with
each other. If we now go to (16), and solve it with the approximation (18),
and we take the angular speed, as given by the Machian expression,

\bigskip

$\omega\cong\frac{\alpha}{r}$
\ \ \ \ \ \ \ \ \ \ \ \ \ \ \ \ \ \ \ \ \ \ \ \ \ \ , ($\alpha\leqslant c=1$), \ \ \ \ \ \ \ \ \ \ \ \ \ \ \ \ \ \ \ \ \ \ \ \ \ \ \ \ \ \ \ \ \ \ \ \ \ \ \ \ \ \ \ \ \ \ \ \ \ \ \ \ (19)

\bigskip

\bigskip where $\alpha$\ \ is a constant that must be found later, and
\bigskip then, the total spin of the rotating Universe will be proportional to
$r^{2}$ \ , i.e.:

\bigskip

$L\cong Mr^{2}\omega=\gamma\alpha r^{2}\propto r^{2}$
\ \ \ \ \ \ \ \ \ \ \ \ . \ \ \ \ \ \ \ \ \ \ \ \ \ \ \ \ \ \ \ \ \ \ \ \ \ \ \ \ \ \ \ \ \ \ \ \ \ \ \ \ \ \ \ \ \ \ \ \ \ \ \ \ \ \ \ \ \ \ \ (20)

\bigskip

We see, that from (16), after some algebra, if \ $\gamma\cong2$\ ,\ then
\ $\alpha\approx c$\ \ , and thus, we make contact with Berman models [3]-[8].

\bigskip

We notice that the local effect at each point of space, makes the total energy
density equal to zero; it is given, in the electromagnetic case, by a
Poynting-like electric field density,

$\bigskip$

$\rho_{TOT}=E_{TOT}^{2}/8\pi\cong0$\ \ \ \ \ \ \ \ \ \ \ \ \ \ \ \ \ \ \ \ . \ \ \ \ \ \ \ \ \ \ \ \ \ \ \ \ \ \ \ \ \ \ \ \ \ \ \ \ \ \ \ \ \ \ \ \ \ \ \ \ \ \ \ \ \ \ \ \ \ \ \ \ \ \ \ \ \ \ \ \ (22)

\bigskip

When we sum for all points of space, we obviously find that the total energy
of the Universe is zero-valued.

\bigskip

{\large \bigskip Section III - Graneau and Graneau's Theory}

Graneau and Graneau[2], discuss a version of inertia theory, that would be
originated in a Machian Newtonian theory. According to those authors, it is an
instantaneous action-at-a-distance theory. The consequence of such theory, is
the "new"\bigskip\ force law for inertia,

\bigskip

$\Delta F_{i}=-\frac{a}{\pi^{2}B}\left[  \frac{m_{0}m_{x}}{r^{2}}\right]  $
\ \ \ \ \ \ \ \ \ \ \ \ \ \ \ \ \ \ \ , \ \ \ \ \ \ \ \ \ \ \ \ \ \ \ \ \ \ \ \ \ \ \ \ \ \ \ \ \ \ \ \ \ \ \ \ \ \ \ \ \ \ \ \ (23)

\bigskip

where, \ $\Delta F_{i}$\ \ is the inertial force between two particles with
masses \ \ $m_{0}$\ \ \ and \ $m_{x}$\ , separated by a radial distance\ \ $r$%
\ \ \ , while \ \ $B$\ \ is a universal constant, relating \ $\Delta F_{i}%
$\ \ with a universal relative acceleration\ \ $a$\ \ between the two objects.

\bigskip

As I have shown elsewhere, Mach's principle can be thought of, as the
mathematical representation of the zero-total-energy of the Machian Universe.
In another paper, Berman[4] calculated the existence of an anomalous universal
acceleration which acts relative to each pair of "observer" and "observed"
objects, due to the same Machian principle. The numerical value of the
expected relative acceleration, acting in a radial direction from the
"observed" to the "observer", is about \ $8.0$ $\times$ $10^{-8}cm/\sec^{2}%
$\ . This result was shown by Berman to agree and explain the so-called
Pioneers' anomalous acceleration relative to the Earth, which is affecting
both spaceships that travel in the outskirts of the Solar system, in two
opposite directions relative to the Earth or approximately towards the Sun.

\bigskip

Graneau and Graneau[2], have failed to determine the numerical value of their
constant \ \ $B$\ \ . By comparing with Newtonian law of gravitation, we may
write, on the assumption that the acceleration to be met is the Pioneers'
anomalous one,

\bigskip

\bigskip$\Delta F_{i}=-\frac{a}{\pi^{2}B}\left[  \frac{m_{0}m_{x}}{r^{2}%
}\right]  =-G\left[  \frac{m_{0}m_{x}}{r^{2}}\right]  $
\ \ \ \ \ \ \ \ \ \ \ \ \ \ \ \ \ \ \ , \ \ \ \ \ \ \ \ \ \ \ \ \ \ \ \ \ \ \ \ \ \ \ \ \ \ (24)

\bigskip

so that,

\bigskip

$a=\pi^{2}BG$ \ \ \ \ \ \ \ , \ \ \ \ \ \ \ \ \ \ \ \ \ \ \ \ \ \ \ \ \ \ \ \ \ \ \ \ \ \ \ \ \ \ \ \ \ \ \ \ \ \ \ \ \ \ \ \ \ \ \ \ \ \ \ \ \ \ \ \ \ \ \ \ \ \ \ \ \ \ \ \ \ \ \ (25)

\bigskip

where, \ $G$\ \ stands for Newton's gravitational constant.

\bigskip

When the numerical values above are plugged, we find:

\bigskip

$B\cong1.0$ \ $m^{-2}kg^{-1}$ \ \ \ \ \ \ \ \ \ \ \ \ \ \ \ . \ \ \ \ \ \ \ \ \ \ \ \ \ \ \ \ \ \ \ \ \ \ \ \ \ \ \ \ \ \ \ \ \ \ \ \ \ \ \ \ \ \ \ \ \ \ \ \ \ \ \ \ \ (26)

\bigskip

This constant is the result of interactions $B_{i}$\ \ from all other masses
in the Universe, in the treatment of Graneau and Graneau, where the \ $B_{i}%
$\ 's are determined by the mass of each "cause"\ , i.e., each mass in the
Universe, divided by its distance to the given local point of space where they
cause the inertia force.\ \ 

\bigskip

Though we have found the otherwise undetermined numerical value for \ $B$\ \ ,
we point out that we do not agree with Graneau and Graneau, when they discard
General Relativity in favor of Newtonian gravitation. In fact, it has been
shown earlier, that cosmological models obeying
Brans-Dicke-Whitrow-Randall-Sciama \ relation,

\bigskip

$\frac{GM}{c^{2}R}\cong1$ \ \ \ \ \ \ \ \ \ \ \ \ \ , \ \ \ \ \ \ \ \ \ \ \ \ \ \ \ \ \ \ \ \ \ \ \ \ \ \ \ \ \ \ \ \ \ \ \ \ \ \ \ \ \ \ \ \ \ \ \ \ \ \ \ \ \ \ \ \ \ \ \ \ \ \ \ \ \ \ \ \ \ \ (26)

\bigskip

which is derived from the Machian zero-total-energy hypothesis, and which
models are based on General Relativity theory or \ alternative generalizations
of such theory, should be regarded as fulfilling the Machian property.

\bigskip

\bigskip{\large \bigskip Section IV - On Berry's Machian Argument }

Berry [9] has posed a Machian query. Consider a body of mass \ $m$\ \ , acted
on by a large one \ $M$\ \ located at a distance \ $r$\ \ , while the large
mass has an acceleration \ $\vec{a}$\ \ relative to the small one. In order to
satisfy Mach's principle, the force exerted on the small mass by the larger,
must contain a part proportional to \ $m$ $\vec{a}$\ \ \ . By means of
dimensional analysis, \ we find that the correct force should be proportional
to \ $m$ $\vec{a}$\ \ \ and also to other terms: \ $M$\ , \ $r$\ \ \ ,
\ \ $G$\ \ \ \ and \ \ $c$\ \ , at some powers. According to Newton's third
law, the power of \ $M$\ \ and \ \ $m$\ \ \ must be the same, i.e., they occur
symmetrically in the force equation. The solution is,

\bigskip

$\vec{F}=-GM$ $\frac{m\text{ }}{c^{2}r}$ $\vec{a}$ \ \ \ \ \ \ \ \ \ \ \ \ \ . \ \ \ \ \ \ \ \ \ \ \ \ \ \ \ \ \ \ \ \ \ \ \ \ \ \ \ \ \ \ \ \ \ \ \ \ \ \ \ \ \ \ \ \ \ \ \ \ \ \ \ \ \ \ \ \ \ \ \ \ \ \ \ \ \ (28)

\bigskip

\bigskip This looks like the force whose acceleration measures mutually
accelerated charges. For the gravitational case, Sciama has given a name to
it: law of inertial induction. For one thing, we may understand from the
analogy, that if electromagnetic radiation is possible, then we would also
have gravitational radiation.

\bigskip

If law (28) is to be applied to the distant masses of the Universe, in the
Machian picture, and if Newton's second law should be valid, we need the
following Brans-Dicke relation to be valid:

\bigskip

$\bigskip\frac{GM}{c^{2}r}=1$ \ \ \ \ \ \ \ \ \ \ \ \ \ \ \ \ \ \ \ \ \ \ , \ \ \ \ \ \ \ \ \ \ \ \ \ \ \ \ \ \ \ \ \ \ \ \ \ \ \ \ \ \ \ \ \ \ \ \ \ \ \ \ \ \ \ \ \ \ \ \ \ \ \ \ \ \ \ \ \ \ \ \ \ \ \ \ \ \ \ \ \ \ \ \ \ \ \ \ \ \ (29)

\bigskip

so that,

\bigskip

$\vec{F}=-GM$ $\frac{m\text{ }}{c^{2}r}$ $\vec{a}=-$ $m$ $\vec{a}%
$\ \ \ \ \ \ \ \ \ \ \ \ \ . \ \ \ \ \ \ \ \ \ \ \ \ \ \ \ \ \ \ \ \ \ \ \ \ \ \ \ \ \ \ \ \ \ \ \ \ \ \ \ \ \ \ \ \ \ \ \ \ \ \ \ \ \ \ \ \ \ \ \ \ \ \ \ \ \ (30)

\bigskip

This section was a digression on a Berry's argument. It must be said that from
formula (30), we have the same kind of zero-total force applied to each and
all particles in the Universe, (inertial force plus gravitational force,
equals zero) so that we retrieve a zero-total energy of the Universe.\bigskip

{\large \bigskip Section V - Conclusions}

Sciama's linear theory, was generalised by including in the same model a
rotating and at the same time expanding Universe. The resultant equations, are
equivalent to the Machian treatment by Berman[3]-[8].

\bigskip

We have also found the constant \ \ $B$\ \ numerical value, but, nevertheless,
we guess that we need not exclude other theories in favor of Newton's one.
Section IV shows the deep value of Machian ideas. Indeed, inertia has been
treated as Machian-originated.

\bigskip\bigskip

\bigskip{\Large Acknowledgements}

\bigskip

I, gratefully, recognize the very important contribution of an anonymous
referee, from whose report I borrowed freely several paragraphs; I also thank
my intellectual mentors, Fernando de Mello Gomide and the recently deceased M.
M. Som, to whom I dedicate this paper; and I am also grateful for the
encouragement by Paula, Albert, and Geni and also to Marcelo F. Guimar\~{a}es,
Nelson Suga, Mauro Tonasse and Antonio F. da F. Teixeira.

\bigskip

{\Large References}

\bigskip

[1]. Sciama, D.W. (1953) - M.N.R.A.S., \textbf{113}, 34.

[2]. Graneau, P.; Graneau, N. (2006) - \textit{In the Grip of the Distant
Universe - the Science of Inertia }, World Scientific, Singapore, pages 256-267.

[3]. Berman, M.S. (2007) - \textit{Introduction to General Relativity and the
Cosmological Constant Problem, }Nova Science, New York.

[4]. Berman, M.S. (2007a) - \textit{Introduction to General Relativistic and
Scalar-Tensor Cosmologies, }Nova Science, New York.

\bigskip\lbrack5]. Berman, M.S. (2007b) - \textit{The Pioneer Anomaly and a
Machian Universe} - Astrophysics and Space Science, \textbf{312},275.

[6]. Berman, M.S. (2008) - \textit{A General Relativistic Rotating
Evolutionary Universe,} Astrophysics and Space Science, \textbf{314, }319-321.

[7]. Berman, M.S. (2008a) - \textit{A General Relativistic Rotating
Evolutionary Universe - Part II,} Astrophysics and Space Science, \textbf{315,
}367-369.

[8]. Berman, M.S. (2008b) - \textit{A Primer in Black Holes, Mach's Principle
and Gravitational Energy, }Nova Science, New York.

[9]. Berry, M.V. (1989) - \textit{Principles of Cosmology and Gravitation,
}Adam Hilger, Bristol.

\end{document}